\documentclass[%
reprint,
 amsmath,amssymb,
 aps,
prl,
]{revtex4-1}

\usepackage{gensymb}
\usepackage{graphicx}% Include figure files
\usepackage{dcolumn}% Align table columns on decimal point
\usepackage{bm}% bold math
\usepackage{physics}
\usepackage{amsmath}
\usepackage{siunitx}
\begin{document}

\preprint{APS/123-QED}

\title{ Frequency-comb based double-quantum two-dimensional coherent spectroscopy identifies collective hyperfine resonances in atomic vapor induced by dipole-dipole interactions}
%Manuscript Title:\\with Forced Linebreak}% Force line breaks with \\

\author{Bachana Lomsadze}
\author{Steven T. Cundiff}%
 \email{cundiff@umich.edu}
\affiliation{%
Department of Physics, University of Michigan, Ann Arbor, Michigan 48109, USA} %\textbackslash\textbackslash
%}%

\date{\today}% It is always \today, today,

\begin{abstract}
Frequency comb based multidimensional coherent spectroscopy is a novel optical method that enables high resolution measurement in a short acquisition time. The method's resolution makes multidimensional coherent spectroscopy relevant for atomic systems that have narrow resonances. We use double-quantum multidimensional coherent spectroscopy to reveal collective hyperfine resonances in rubidium vapor at $100 \degree $C induced by dipole-dipole interactions. We observe tilted lineshapes in the double-quantum 2D spectra, which has never been reported for Doppler-broadened systems. The tilted lineshapes suggest that the signal is predominately from the interacting atoms that have near zero relative velocity. 

\begin{description}
\item[PACS numbers]
34.20.Cf, 42.62.Fi, 78.47.nj %May be entered using the \verb+\pacs{#1}+ command.
\end{description}
\end{abstract}

\pacs{Valid PACS appear here}% PACS, the Physics and Astronomy
\maketitle

%\tableofcontents

%\section{\label{sec:level1}First-level heading:\protect\\ The line
%break was forced \lowercase{via} \textbackslash\textbackslash}

Dipole-dipole interactions are among the most fundamental and important processes in atomic, molecular and optical physics. Understanding these interactions are crucial because they govern the physical mechanisms of many phenomena. Dipole-dipole interactions result in energy transfer between atoms, molecules and complex biological systems \cite{energy_transfer,RN59,RN55}. They play the major role for formation of homo and hetero-nuclear and exotic molecules~\cite{RN60}. These interactions are also critical for many applications such as quantum computing, Rydberg blockades and designing single quantum emitters~\cite{Blockade, RN61, qua_com}.  

Since its development over two decades ago, optical multidimensional coherent spectroscopy (MDCS) \cite{stevemukamel, hamm2011concepts} has proven to be a powerful optical method for probing weak many-body interactions. It is an optical analog of multidimensional nuclear magnetic resonance spectroscopy \cite{NMR} that has been a workhorse for several decades for determining the molecular structure. Optical MDCS is a non-linear technique that uses a sequence of ultrafast laser pulses (typically three) incident to the sample and records a non-linear (four-wave-mixing (FWM)) signal emitted by the sample as a function of the time delay(s) between the incident pulses. A multidimensional spectrum is constructed by calculating the Fourier transforms of the emitted signal with respect to the emission time and the delays between the pulses. Depending on the time ordering of the excitation pulses, a multidimensional spectrum can give insight about many-body interactions  and provide important spectroscopic information. For instance, if the  photon-echo excitation sequence \cite{photonecho} is used, when the first pulse is a complex phase-conjugated pulse, a multidimensional spectrum (referred to as a single-quantum 2D spectrum) shows the couplings between the excited states, and it also differentiates the homogenous and inhomogeneous linewidths. Single-quantum spectra can also be used for chemical sensing applications to determine the constituent species in a mixture \cite{Lomsadze1389}. If the complex conjugated pulse arrives last then the corresponding 2D spectrum (referred to as a double-quantum spectrum) can identify weak many-body interactions \cite{Yang:mukame1,Yang:mukamel2}. Until this point however, due to the resolution and acquisition-speed limitations, MDCS techniques have mostly been used for systems that have broad resonances or fast dephasing rates (tens of fs to hundreds of ps). They have not been able to probe fundamental processes such as the dipole-dipole interactions in atomic systems (with nanosecond dephasing times) that are the building blocks for complex matter. 

Previously, single and double-quantum MDCS measurements have been applied to Rubidium (Rb) and Potassium (K) atomic vapors (at 130\textsuperscript{o}C) to investigate collective resonances induced by weak dipole-dipole interactions \cite{Gao:16, Dai}. However, due to limited spectrometer resolution, an Argon (Ar) buffer gas was introduced into the vapor cell to artificially broaden the resonances to match the spectrometer resolution. The broadening led to the modification (distortion) of the natural Doppler-broadened line shapes. It is important to emphasize that obtaining undistorted line shapes is extremely critical as the lineshapes provide insight about the underlying physics of the many-body interactions. In addition, the experimental measurements \cite{Gao:16, Dai} could not differentiate homo-nuclear (between same isotopes) and hetero-nuclear (between different isotopes) interactions. Collective resonances in a dilute Potassium vapor were also studied by L. Brunder et al. \cite{Brunder} and theoretically explained by S. Mukamel \cite{Mukamel:non:interact}, however the detected signal was due to non-interacting atoms and hence contained no information about the dipole-dipole interactions.    % which is extremely important for chemical sensing applications.  

\begin{figure*}
\includegraphics[width=0.9\textwidth]{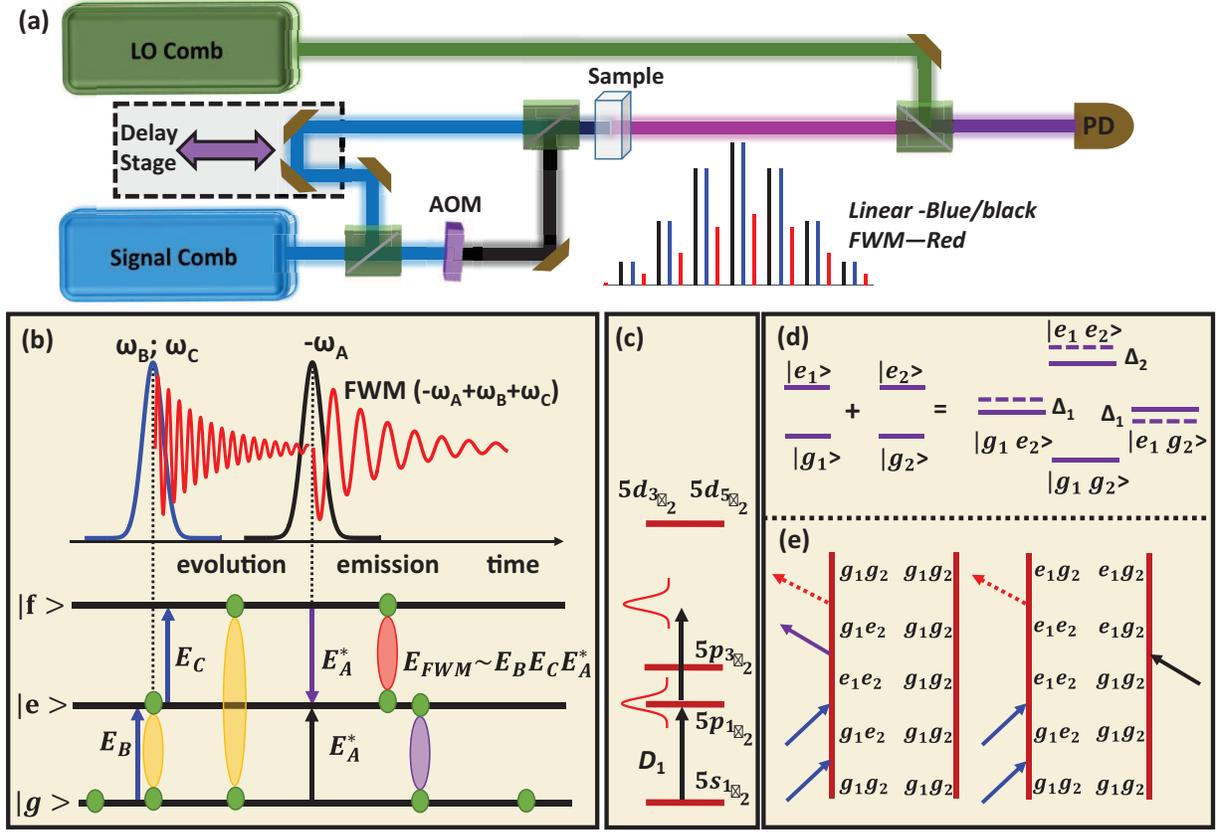}
\caption{\label{fig:exp} (a) Schematic diagram of the experimental setup. AOM Acousto-optical modulator. PD - photodetector. Comb structure shown corresponds to linear (blue and black) and four-wave-mixing (red) comb lines in the frequency domain. (b) time domain picture of FWM signal generation. $\ket{g}$, $\ket{e}$ and $\ket{f}$ correspond to ground, excited and doubly excited states, respectively.    (c) Fine structure  of Rb atoms, showing no energy level at $2\times D_1$ frequency.  (d) energy level diagram of 2 combined atoms without interactions. Dashed lines show the energy levels with interactions. (e) Double-sided Feynman diagrams of the double-quantum FWM signals.}
\end{figure*}

Recently, we introduced a novel approach \cite{Lomsadze1389} to multidimensional coherent spectroscopy that utilizes frequency combs and the dual-comb detection technique \cite{Jun_steve, Coddington:16}. This combination allowed us to demonstrate rapid single-quantum two-dimensional coherent spectroscopy with unprecedented resolution (hundreds of MHz)~\cite{Lomsadze1389}. Here, we take advantage of the speed and resolution achievable with the technique and extend its applications to double-quantum MDCS, investigating dipole-dipole interactions in atomic vapor. We apply our method to a vapor of Rb atoms containing both isotopes \textsuperscript{87}Rb and \textsuperscript{85}Rb at their natural abundance with Doppler-broadened features (at $100 \degree$C) and observe  collective hyperfine resonances (both homo-nuclear and hetero-nuclear) induced by weak dipole-dipole interactions. Our results also reveal that the FWM signal, due to many-body interactions, is stronger for the atoms that have near zero relative velocity.

The experimental setup is shown in Fig. \ref{fig:exp} (a) with further details available in Refs. \onlinecite{Lomsadze1389,Lomsadze:17}.  We used two home-built Kerr-lens mode-locked Ti:Sapphire  lasers centered at 800 nm. The repetition frequencies for the signal and the LO combs (f\textsubscript{rep\textsubscript{-sig}}=93.581904 MHz and f\textsubscript{rep\textsubscript{-LO}}=f\textsubscript{rep\textsubscript{-sig}} - 641 Hz)  were phase-locked to a direct digital synthesizer, but the comb offset frequencies were not actively stabilized. The phase fluctuations due to fluctuations in offset frequency, optical path length and/or repetition frequency were measured and corrected using a scheme described in \cite{Lomsadze1389,Lomsadze:17}, which is similar to the phase correction schemes that are used in linear dual-comb spectroscopy \cite{Roy:12,Ideguchi:12,ideguchi2014adaptive}. The output of the signal comb was split into 2 parts. One part of the beam was frequency shifted by 80 MHz using an acousto-optical modulator and combined with the other part whose delay was controlled with the retro-reflector mounted on a mechanical stage. The combined beams then were focused to 5 \SI{}{\micro\metre} spot in a 0.5 mm thin vapor cell containing \textsuperscript{87}Rb and \textsuperscript{85}Rb atoms (at $100 \degree$C). Average powers per beam were 2.4 mW and 1.2 mW respectively.  Before focusing, the beams were filtered with an optical bandpass filter centered at 794 nm (3 nm FWHM) to excite only the $D_1$ lines of both isotopes.  The generated FWM signal comb, along with the excitation combs, were then combined with the LO comb with slightly different repetition rate, and interfered on a photodetector. The output of the photodetector was spectrally filtered in the RF domain to isolate the FWM signal and digitized \cite{Lomsadze:17}. The delay between the excitation pulses was varied to generate the second dimension for the double-quantum two-dimensional spectrum.  

\begin{figure}[t]
\includegraphics[width=0.53\textwidth]{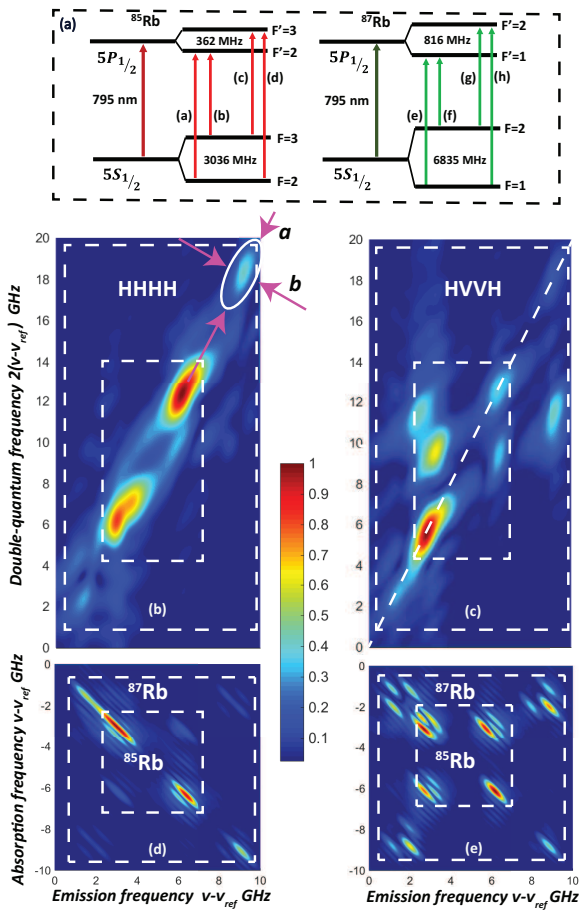}
\caption{\label{fig:results} (a) Energy level diagrams of $D_1$ hyperfine lines of \textsuperscript{87}Rb and \textsuperscript{85}Rb atoms. (b) and (c) double-quantum two-dimensional spectra acquired by co-linearly and cross-linearly polarized excitation pulses. H-horizontal, V-vertical. (d) and (e) single quantum two-dimensional spectra. Color scale shows normalized signal magnitude.}
\end{figure}

The generation of a double-quantum FWM signal in the time domain with a pair of pulses is pictorially shown in Fig. \ref{fig:exp} (b). The first pulse (shown in blue) excites a coherence between the ground and singly excited states and then converts this coherence into the double-quantum coherence between the ground state and doubly-excited state that evolves in time (red trace in the figure shows the evolution of the coherence in time) \cite{endnote_1}. Pulse (A) (complex-conjugate pulse shown in black) converts this coherence either back  to the coherence between the ground and the excited state or to the coherence between the excited and doubly-excited states that radiates the FWM signal (red trace in the figure). As mentioned earlier, the excitation beams in our experiment were optically filtered to excite only $D_1$ lines of Rb atoms and there are no doubly-excited states in Rb within the filtered bandwidth at $2\times D_1$ frequency (see Fig. \ref{fig:exp} (c)). In this case the only way to obtain the double-quantum FWM signal is to consider a combined atom picture that clearly shows the doubly-excited state (Fig. \ref{fig:exp} (d)). In Fig. \ref{fig:exp} (e), we plot the double-sided Feynman diagrams \cite{Feynman} for the combined atom picture that would give rise to the FWM signal. However, the Feynman diagrams have opposite signs and since $\ket{g_{1} e_{2}} - \ket{g_{1}g_{2}}$ and  $\ket{e_{1}e_{2}} -\ket{e_{1}g_{2}}$   transition energies are equal to each other, these double-sided Feynman diagrams perfectly cancel each other.  The picture changes if we include the many-body interactions, particularly the dipole-dipole interactions  \cite{Yang:mukame1,Yang:mukamel2}. In the presence of the interactions  the singly and doubly-excited states experience slight energy shifts (dashed lines in Fig. \ref{fig:exp} (d) ) or changes in linewidth. These effects ($\Delta_{1}$ and $\Delta_{2}$) are enough to break the symmetry between the states and lead to generation of a FWM signal.

In Fig.~\ref{fig:results} we show the results. Figure~\ref{fig:results} (a) shows the $D_1$ hyperfine states of both isotopes. Fig.~\ref{fig:results} (b) and (c) correspond to double-quantum 2-dimensional spectra obtained with co-linearly (HHHH) and cross-linearly (HVVH) polarized excitation pulses. The diagonal peaks (along the line from (0, 0) GHz to (10, 20) GHz) correspond to coupling between the same hyperfine energy levels of two atoms of the same isotopes (outer white dashed box for \textsuperscript{87}Rb) and (inner white dashed box for \textsuperscript{85}Rb). The off-diagonal peaks show coupling between different hyperfine energy levels of two atoms of the same as well as  different isotopes.  For instance, in Fig.~\ref{fig:results} (c) the peak at (9.0, 18.0) GHz corresponds to the coupling of two \textsuperscript{87}Rb atoms that have the same (h) hyperfine resonance frequencies, whereas the peaks around (9.0, 11.2) GHz and (2.2, 11.2) corresponds to coupling of two \textsuperscript{87}Rb atoms with (h) and (g) hyperfine resonance frequencies, respectively. The peaks at (1.3, 4.2) GHz and (3.0 4.2) corresponds to the coupling of \textsuperscript{87}Rb and \textsuperscript{85}Rb isotopes with (f) and (c) resonance frequencies, respectively. The similar analysis can be performed to identify all the peaks in double-quantum 2D spectra.

It is important to emphasize that double-quantum MDCS excels in isolating and identifying many-body interactions because it allows the measurement of the FWM signal that is only due to the interactions. These interactions are, in most cases, not accessible with other methods, including single-quantum MDCS. To demonstrate this point, we compared double-quantum 2D spectra to single-quantum 2D spectra shown in Fig.~\ref{fig:results} (d) and (e)  (taken by co-linearly and cross-linearly polarized excitation pulses, respectively). The spectra were taken with the pulse ordering that leads to formation of a photon echo (the complex conjugated pulse arrives first), which can be experimentally obtained by swapping the time order of the excitation pulses such that the AOM frequency shifted pulse (A) arrives first (Fig. \ref{fig:exp} (a)). The diagonal elements (along (0,0) GHz and (10, -10) GHz line) correspond to FWM signals with the same absorption and emission hyperfine frequencies (a-h) for \textsuperscript{87}Rb (outer white dashed box) and \textsuperscript{85}Rb (inner white dashed box). They are diagonally elongated due to Doppler broadening. The cross-peaks, on the other hand, show all possible couplings between the hyperfine states within the same atom. In the photon echo excitation sequence the FWM signal due to the couplings of 2 different atoms via the dipole-dipole interaction is non-zero. However due to its weak strength compared to the FWM signal from individual atoms, the coupling peaks are not visible on 2D spectra. This shows that the single-quantum MDCS is not sensitive enough to probe the weak many-body interactions in atomic/molecular systems and measuring double-quantum spectra is required to isolate these interactions. 

The double-quantum spectra show additional interesting behavior. The peaks are tilted along the diagonal line. The tilted peaks (along the diagonal) are expected for single-quantum spectra because the pulses' time ordering produces a photon echo scheme. Double-quantum spectra, on the other hand, use a pulse time ordering that should not lead to photon echo. The tilted line shapes on double quantum spectra have previously been observed in molecules \cite{muka_elongation, mol_elongation} and in static inhomogeneously broadened semiconductor materials \cite{Tollerud:16} but have never been reported for Doppler-broadened systems. The elongation suggests that there is a correlation between the emission and double-quantum frequencies that gives insight about what velocity group of atoms participate in generation of the FWM signal. 

To give quantitative information we measured the correlation parameter

\begin{equation}
    C = \frac{a^{2}-b^{2}}{a^{2}+b^{2}}
\end{equation}
where $a$  and $b$ are the sizes of the ellipse along the major and minor axe, shown in Fig.~\ref{fig:results}(b) (upper right corner). $C=0$ would imply that any two atoms couple to each other via the dipole-dipole interaction and  give rise to the FWM signal. Whereas $C=1$ would indicate that the FWM signal is predominantly from two interacting atoms that have near zero relative velocity. We chose the isolated peak that corresponds to coupling of two \textsuperscript{87}Rb atoms with (h) resonance frequencies and we measured the correlation to be 0.85. Even higher correlation can be obtained by improving the signal to noise ratio. This value is a very high correlation and indicates that the FWM signal is due to the coupled atoms with near zero (within 15 \%) relative velocities. A plausible explanation of the high correlation could be the fact that the dipole-dipole interaction is proportional to (1/r\textsuperscript{3}), where $r$ is the inter nuclear separation between the atoms. If two atoms have non-zero relative velocity then their inter-nuclear separation changes during the time between second and third excitation pulses (that is scanned up to 1 ns). For high relative velocities this could causes the dipole-dipole interaction to degrade rapidly (1/r\textsuperscript{3})  and hence to decrease the strength of the FWM signal. 

In conclusion, we have demonstrated the measurement of collective hyperfine resonances in a vapor of Rb atoms induced by the dipole-dipole interactions. We have identified the peaks corresponding to the couplings between the hyperfine levels of two atoms of the same and different isotopes. We have reported tilted peaks in double-quantum 2D coherent spectra for Doppler broadened system and provided the quantitative information about the velocity groups of the atoms that participate in dipole-dipole interactions. %We have also showed the potential application of single and double quantum high resolution spectra in chemical sensing.

The combination of single and double-quantum spectra makes frequency comb-based multidimensional coherent spectroscopy extremely powerful tool for obtaining complete spectroscopic information about atomic and molecular systems.  

The research is based upon work supported by the Office of the Director of National Intelligence (ODNI), Intelligence Advanced Research Projects Activity (IARPA), via contract 2016-16041300005. The views and conclusions contained herein are those of the authors and should not be interpreted as necessarily representing the official policies or endorsements, either expressed or implied, of the ODNI, IARPA, or the U.S. Government. The U.S. Government is authorized to reproduce and distribute reprints for Governmental purposes notwithstanding any copyright annotation thereon.

We thank Christopher Smallwood and Eric Martin for helpful discussions.

\bibliography{references}% Produces the bibliography via BibTeX.

\begin{thebibliography}{29}
\bibitem [{end()}]{endnote_1}%
  \BibitemOpen
  \href@noop {} {}\bibinfo {note} {For clarity we separated the first and
  second interactions but in the experiment these interactions happen
  simultaneously}\BibitemShut {NoStop}%
\end{thebibliography}

\end{document}